\documentclass{vietnam}

\def\Journal#1#2#3#4{{#1} {\bf #2}, #3 (#4)}

\def\NPB{{\em Nucl. Phys.} B}

\def\PRD{{\em Phys. Rev.} D}

\def\be{\begin{equation}}
\def\ee{\end{equation}}
\def\bea{\begin{eqnarray}}
\def\eea{\end{eqnarray}}

\graphicspath{ {figures/} }

\newcommand{\Photo}{}

\begin{document}
\vspace*{4cm}
\title{Constraints on large scalar multiplets added to the Standard Model}

\author{ Darius~Jur\v{c}iukonis$^{(1)}$ and Lu\'\i s~Lavoura$^{(2)}$ }

\address{ $^{(1)}\!$
  \small Vilnius University, Institute of Theoretical Physics and Astronomy, \\
  \small Saul\.etekio~av.~3, Vilnius 10257, Lithuania
  \\*[2mm]
  $^{(2)}\!$
  \small Universidade de Lisboa, Instituto Superior T\'ecnico, CFTP, \\
  \small Av.~Rovisco~Pais~1, 1049-001~Lisboa, Portugal}

\maketitle\abstracts{
We study the extension of the Standard Model (SM) by introducing a scalar multiplet with arbitrary isospin $J$ and hypercharge $Y$. 
We explicitly consider various possible values of the weak isospin $J$, up to and including $J=7/2$. 
The mass differences among the components of the multiplet originate from its coupling to the Higgs doublet of the SM, as present in the scalar potential (SP). We derive exact bounded-from-below (BFB) and unitarity (UNI) conditions for this model, even when the SP includes the most general quartic terms involving the multiplet components.
We find that the upper bound on the mass differences depends not only on the UNI conditions but also on the BFB ones, thus imposing constraints on the mass differences. 
We compare these constraints to those derived from the oblique parameters (OPs) and from solutions of the renormalization-group equations (RGEs).
}

\section{Introduction}

This paper is based on Ref.~\cite{multiplets}, 
where we study the model in which one adds to SM
a gauge-$SU(2)$ multiplet $\chi$
with weak isospin $J$,
consisting of $n = 2 J + 1$ complex scalar fields.
It is assumed that the scalar fields 
do not have any vacuum expectation value (VEV), while
the multiplet has unspecified weak hypercharge $Y$.

In the SP, there is a renormalizable coupling
\be
\lambda_4\, \sum_{a=1}^3 \left( H^\dagger\, \frac{\tau_a}{2}\, H \right)
\left[ \chi^\dagger\, T_a^{(J)}\, \chi \right]
\label{coup}
\ee
of $\chi$ to the Higgs doublet $H$ of the SM, 
where $\lambda_4$ is a dimensionless coefficient, 
the $\tau_a$ are the Pauli matrices, 
$\chi$ is a column vector of $n$ scalar fields, 
and the $T_a^{(J)}$ are the $n \times n$ matrices  that represent $su(2)$ in the $J$-isospin representation.
This coupling generates
a fixed (constant) squared-mass difference, $\Delta m^2 \propto v^2$
($v$ is the VEV of $H$),
between any two components of $\chi$ whose third component of isospin
differs by one unit.

This model was first considered in Ref.~\cite{li}
as a paradigm for potentially large oblique parameters.
In Refs.~\cite{logan,milagre}, mixed constraints on $J$ and $Y$ were derived, showing that $n$ cannot exceed eight (\textit{i.e.} $J \le 7/2$) due to perturbative unitarity restrictions.
Other recent papers
that consider scalar quadruplets with specific hypercharges
are Refs.~\cite{kannikerecent,jurc}.

In this analysis, we constrain the modulus of the coefficient $\lambda_4$
of the term in Eq.~\ref{coup} by considering both the UNI and BFB 
conditions on the quartic part of the SP.
Additionally, we consider the confrontation of our model
with the RGEs and OPs that it generates.
These constraints place an upper bound on $\Delta m^2$.

\section{The scalar potential without terms four-linear on $\chi$}

The scalar potential of the model is given by
\be
V = \sum_{k=1}^2 \left( \mu_k^2 F_k + \frac{\lambda_k}{2}\, F_k^2 \right)
+ \lambda_3 F_1 F_2 + \lambda_4 F_4
+ \sum_{i=5}^{t+3} \lambda_i F_i,
\label{potentialV}
\ee
but in this section we set $\lambda_2 = 0$ and discard all the other terms four-linear in the $\chi_I$, \textit{i.e.}\ the last term in the potential $V$. Here, $t=\mathrm{ceil} (n/2)$ is the ceiling function, and $I$ is the third component of weak isospin. 
Furthermore, $F_1 = |H_{1/2}|^2 + |H_{-1/2}|^2$ and $F_2 = \sum_{I=-J}^{I=J} |\chi_I|^2$.
Since experimentally $m_H \approx$ 125\,GeV and $v \approx$ 174\,GeV,
we assume that $\lambda_1 \approx 0.258$.

From the potential in Eq.~\ref{potentialV}, we can derive the mass-squared $m_I^2$ of the scalar $\chi_I$, which implies that the difference between the squared masses $\Delta m^2$
of $\chi_I$ and $\chi_{I+1}$ is
\be
m_I^2 = \mu_2^2 + \left( \lambda_3 - \frac{\lambda_4}{2}\, I \right)
\left| v \right|^2 
\quad \Longrightarrow \quad	
\Delta m^2 = \frac{\left| \lambda_4 v^2 \right|}{2},
\label{deltam}
\ee
which is independent of $I$.
The upper bound on $\left| \lambda_4 \right|$
is
equivalent to an upper bound on $\Delta m^2$.

\paragraph{UNI and BFB conditions} 
~\\
For UNI conditions, we consider the scattering of a pair of scalars from the multiplet $\chi$
to another pair of scalars,
with both pairs having the same third component of isospin $I$ and hypercharge $Y$.
The necessary and sufficient BFB conditions
for $V_4$ to remain non-negative for arbitrary values of the fields are derived from copositivity conditions~\cite{kannike}.

\begin{figure}[ht]
\begin{center}
\includegraphics[width=1.0\textwidth]{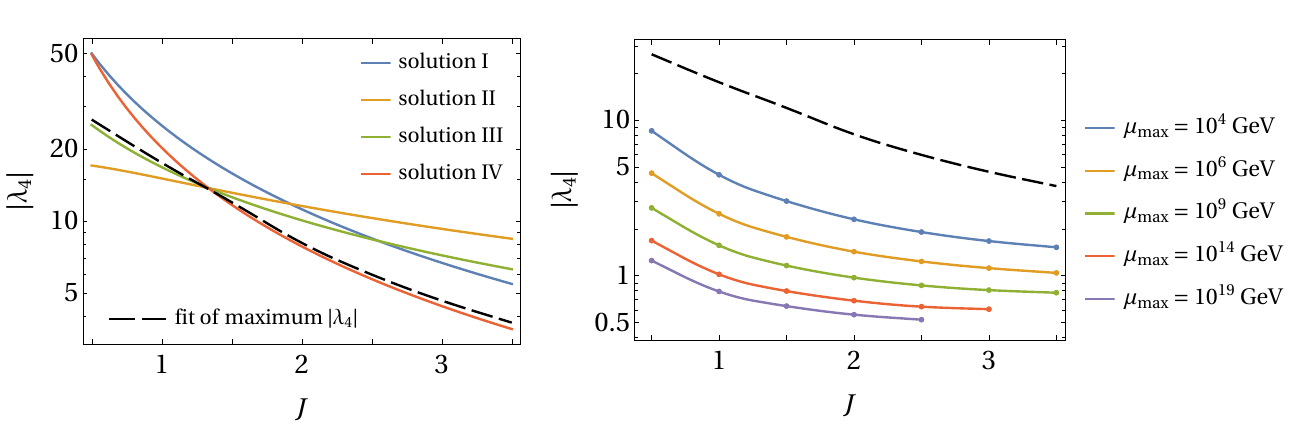}
\end{center}
\vspace{-4mm}
\caption{Left panel: The solutions for $\left| \lambda_4 \right|$ \textit{versus} $J$, with $M = 8 \pi$. 
The black dashed line indicate the maximum allowed values of $\left| \lambda_4 \right|$, as described in Section~\ref{sec:fullPot}. 
Right panel: The maximum allowed values of $\left| \lambda_4 \right|$ \textit{versus} $J$ for different cut-off scales $\mu_\mathrm{max}$. 
The black dashed line are the same as those in the left panel.}
\label{fig:lmd4_solut}
\end{figure}

By transforming all the relevant UNI and BFB
conditions into strict equalities,
we obtain the following equations:
\bea
\lambda_3 + \frac{J + 1}{2} \left| \lambda_4 \right| &=& M,
\label{primeira} \\
\lambda_1 + \sqrt{\lambda_1^2 + \frac{2 J \left( 2 J^2 + 3 J + 1 \right)
  \lambda_4^2}{3}} &=& 2 M,
\label{segunda} \\
3 \lambda_1 + \sqrt{9 \lambda_1^2 + 8 \left( 2 J + 1 \right)
  \lambda_3^2} &=& 2 M,
\label{terceira} \\
\left| \lambda_4 \right| &=& \frac{2}{J}\, \lambda_3,
\label{quarta}
\eea
where $M = 8 \pi$ is the UNI upper bound on $\Re a_0$,
and we have taken into account that both $\lambda_1$ and $\lambda_3$
are non-negative, as required by the BFB conditions. 
These equations can be solved for $\left| \lambda_4 \right|$ and $\lambda_3$.
Equation~\ref{segunda} gives solution I,
equations~\ref{primeira} and~\ref{terceira} together yield solution~II,
equations~\ref{primeira} and~\ref{quarta} lead to solution~III,
and equations~\ref{terceira} and~\ref{quarta} together give solution~IV.
Solutions~I, II, III,
and IV for $\left| \lambda_4 \right|$ 
are plotted in the left panel of Fig.~\ref{fig:lmd4_solut}.

\paragraph{Renormalization-group equations} 
~\\
We consider only the one-loop renormalization group equations
$16 \pi^2 \mu\, \frac{\mathrm{d} g}{\mathrm{d} \mu} = \beta_g$,
where $g$ denotes a generic dimensionless coupling
and $\beta_g$ is a function of
the dimensionless couplings in the theory.
The dimensionless couplings taken into account here are
$g_1$,
$g_2$,
$g_3$,
$\lambda_1$,
$\lambda_2$,
$\lambda_3$,
$\lambda_4$,
and $y_t$.
To derive the RGEs for these eight coupling constants, we used the software {\tt SARAH}~\cite{sarah}.

We solved the differential equations
starting at the scale $\mu = m_t = 173.1$\,GeV 
and let $\mu$ evolve up to the scale
$\mu_\mathrm{Planck} = 10^{19}$\,GeV.
We fix $g_1$, $g_2$, $g_3$, $\lambda_1$ and $y_t$ at $\mu = m_t$ according to Ref.~\cite{huang},
while $\lambda_3$ and $\lambda_4$ are allowed to vary freely, 
subject to the UNI and BFB conditions
at every $\mu$ up to $\mu_\mathrm{max}$.
The right panel of Fig.~\ref{fig:lmd4_solut} shows the results, 
in the form of upper bounds on $\left| \lambda_4 \right|$ 
at the scale $\mu = m_t$, 
depending on the scale $\mu_\mathrm{max}$.
One sees that the upper bound on
$\left| \lambda_4 \right|$ becomes much stricter
when one requires the UNI and BFB conditions
to hold up to $\mu_\mathrm{max}$,
even if $\mu_\mathrm{max}$ is rather low.

The {\tt SARAH} model files, 
as well as the expressions for the RGEs, 
are available in both PDF and {\tt Mathematica} notebook formats at
\href{https://github.com/jurciukonis/RGEs-for-multiplets}{{\tt
    https://github.com/jurciukonis/RGEs-for-multiplets}}.

\section{The scalar potential with terms four-linear in $\chi$}\label{sec:fullPot}

The full potential is given in Eq.~\ref{potentialV}.
The product $\chi \otimes \chi$ of two identical $SU(2)$ multiplets
has only a symmetric component, which consists
of $t$ multiplets
\be
\left( \chi \otimes \chi \right)_\mathrm{symmetric}
= c \oplus d \oplus e \oplus \cdots \oplus q,
\ee
where $c$ is an $SU(2)$ multiplet with weak isospin $2 J$,
$d$ is a multiplet with weak isospin $2J - 2$, 
$e$ has weak isospin $2J - 4$, and so on. 
Lastly, $q$ is either a triplet of $SU(2)$ if $J$ is half-integer, 
or $SU(2)$-invariant if $J$ is integer.
The two-field states in each multiplet
are evaluated using Clebsch--Gordan
coefficients in the standard fashion.

\paragraph{UNI and BFB conditions} 
~\\
There are two main channels that produce UNI conditions:
the two-field states with $I = 0$ and hypercharge $2Y$,
and the two-field states with $I = Y = 0$.
After computing the eigenvalues of the scattering matrices for these states,
all unitarity conditions are determined. 
The explicit expressions for the UNI conditions are provided in Ref.~\cite{multiplets}.

Following the approach in Ref.~\cite{fonseca}, which uses dimensionless quantities in the quartic part of the scalar potential, we derive the necessary BFB conditions:
\vspace{-1mm}
\bea
\lambda_1 &\ge& 0, \quad \lambda_2 \ge 0, \quad \widehat{\lambda}_i \ge 0, \\
\lambda_3 &\ge& - \sqrt{ \lambda_1 \lambda_2}, \quad 
\\
\left| \lambda_4 \right| &\le&
\frac{2}{J} \left( \lambda_3 + \sqrt{\lambda_1 \lambda_2} \right),
\eea
where $\widehat{\lambda}_i \equiv \lambda_2 + q_i$, $q_i \equiv \frac{2 J^2}{\kappa_i}\, \lambda_i$, $\kappa_i = \left( i - 4 \right) \left( 4 J + 9 - 2 i \right)$ and $i = 5, \ldots, t+3$.
Additionally, the sufficient conditions for the boundedness-from-below of $V_4$ are:
\vspace{-2mm}
\bea
&& \mbox{either} \quad \lambda_i > 0, \quad 
\mbox{or} \quad \lambda_i \Lambda_i < 0, 
\\
&& \mbox{or} \quad
\sqrt{\frac{\widehat{\lambda}_i}{q_i\, \Lambda_i}}
> \frac{2}{J \left| \lambda_4 \right|}, \quad
\mbox{or} \quad
\lambda_3 \ge - \sqrt{\frac{\widehat{\lambda}_i\, \Lambda_i}{q_i}}, 
\eea
where $\Lambda_i \equiv \frac{J^2}{4}\, \lambda_4^2 + q_i \lambda_1$.
The maximum allowed value of $\left| \lambda_4 \right|$, 
respecting both the UNI and BFB conditions, 
is illustrated by the black dashed line in Fig.~\ref{fig:lmd4_solut}.

\paragraph{Oblique parameters}  
~\\
We parameterize the corrections to the self-energies of the gauge bosons produced by the new
scalars through six oblique parameters $S$, $T$, $U$, $V$, $W$, and $X$.
We then use these parameters to describe, for each electroweak observable $O$, the ratio between the prediction of the model and that of SM, using expressions of the general form
\be
\frac{O_\mathrm{NP}}{O_\mathrm{SM}} = 1
+ c_S^O S + c_T^O T + c_U^O U + c_V^O V + c_W^O W + c_X^O X, 
\label{ratioNP_SM}
\ee
where the coefficients $c_S^O, \ldots, c_X^O$ are known functions~\cite{dudenas} of the input quantities, while explicit formulas for the OPs are given in Ref.~\cite{multiplets}.

\begin{figure}[ht]
\begin{center}
\includegraphics[width=0.9\textwidth]{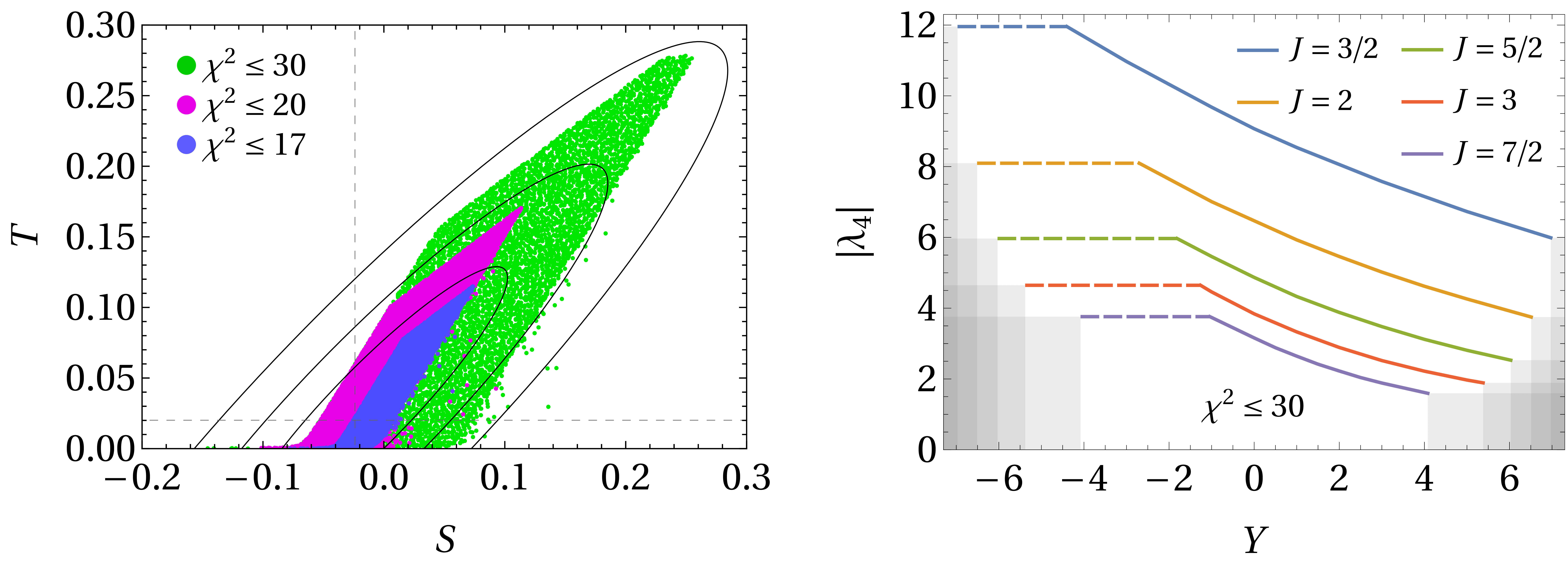}
\end{center}
\vspace{-4mm}
\caption{Left panel: The $S$ and $T$ oblique parameters satisfying specific $\chi^2$ bounds. 
The black ellipses indicate the $1\sigma$, $2\sigma$, and $3\sigma$ bounds of the free $S$-$T$ fit.
Right panel: The upper bound on $\left| \lambda_4 \right|$ \textit{versus} the hypercharge $Y$ 
for various values of $J$, with $m = 3$~TeV and 
fits where $\chi^2 \leq 30$. 
The horizontal dashed lines represent the upper bounds from the UNI + BFB conditions, 
while the curved lines show the upper bounds from the OPs. 
The gray bands indicate the $J$-dependent restrictions on $Y$.
  }
\label{fig:lmd4_OBs}
\end{figure}

For each value of $J$ up to $7/2$, we let $Y$ and 
$\left| \lambda_4 \right|$ vary within the ranges described above, 
while allowing the minimum mass of a multiplet $m$ to vary from 50~GeV to 3~TeV.
We compute the $\chi^2$ function for each set of OPs and
keep only the points with $\chi^2$ smaller than 30 and all the pulls
smaller (in modulus) than three. 
These points lie within the $3 \sigma$ bounds of the free $S$-$T$ fit, as shown in the left panel of Fig.~\ref{fig:lmd4_OBs}.

In the right panel of Fig.~\ref{fig:lmd4_OBs},
which was made for $m = 3$~TeV,
one observes that as $Y$ increases,
the upper bound on $\left| \lambda_4 \right|$ decreases.
It can be seen that unless $m$ is very large 
(and therefore the OPs are very small), 
the constraints on $\left| \lambda_4 \right|$ from the OPs 
are usually stronger than the UNI + BFB conditions considered in this paper.
A full description of the numerical computations is provided in Ref.~\cite{multiplets}.    

\vspace{-4mm}
\section*{Acknowledgments}
D.J.\ research has been carried out in the framework of the agreement of Vilnius University with the Lithuanian Research Council No. VS-13.

\end{document}